\documentclass[journal, onecolumn]{IEEEtran}

\usepackage{cite}
\usepackage{amsmath,amssymb,amsfonts}
\usepackage{algorithmic}
\usepackage{graphicx}
\usepackage{textcomp}
\usepackage{xcolor}
\usepackage{booktabs} 
\usepackage{float}
\usepackage{comment}
\usepackage[caption=false,font=footnotesize]{subfig}
\usepackage[colorlinks=true, citecolor=blue, linkcolor=blue, urlcolor=blue]{hyperref}
\usepackage{orcidlink}

\begin{document}

\title{2D Spatial Indexing for Generalized and Quadrature Spatial Modulation in PASS}
\author{Abanoub Kamel\raisebox{1ex}{\orcidlink{0009-0003-7701-8137}}, \IEEEmembership{Graduate Student Member,~IEEE},  and Ibrahim Al-Nahhal\raisebox{1ex}{\orcidlink{0000-0002-8604-392X}},~\IEEEmembership{Senior Member,~IEEE}
\thanks{The authors are with the Faculty of Engineering and Applied Science, Memorial University of Newfoundland, St. John's, NL A1C 5S7, Canada (e-mail: aehkamel@mun.ca; ioalnahhal@mun.ca).}}

\maketitle

\begin{abstract}
In this letter, three novel spatial modulation (SM) schemes specifically designed for pinching-antenna systems (PASS) are proposed, offering a flexible trade-off between bit error rate (BER) performance and decoder computational complexity. The proposed one-dimensional quadrature SM (1D-QSM) scheme exploits dual waveguides to independently transmit the orthogonal symbol components to eliminate mutual interference. Meanwhile, the 2D-QSM and 2D-generalized SM (2D-GSM) schemes jointly utilize waveguide and pinch indices to enhance spectral efficiency. Results demonstrate that 1D/2D-QSM schemes achieve significantly lower computational complexity and superior BER performance at low spectral efficiencies compared to existing literature and the proposed 2D-GSM. In contrast, 2D-GSM provides enhanced reliability at high spectral efficiencies, outperforming both existing literature and the proposed 1D/2D-QSM schemes. Finally, analytical upper bounds for BER and decoder computational complexity are rigorously derived and validated through simulations.
\end{abstract}

\section{Introduction}

\IEEEPARstart{T}{he} transition to next-generation networks demands unprecedented spectral efficiency and massive data rates \cite{Saad2020}. While conventional massive multiple-input multiple-output (MIMO) systems are foundational, multiplexing data streams across dense arrays induces inter-channel interference (ICI) and necessitates a dedicated radio-frequency (RF) chain per active antenna \cite{DiRenzo2014}. Consequently, signal processing complexity, hardware costs, and power consumption scale prohibitively. Overcoming these challenges requires novel transmission paradigms that boost spectral efficiency without proportionally scaling transceiver complexity.

Spatial modulation (SM) addresses these hardware constraints by completely mitigating ICI. While its predecessor, space shift keying (SSK), conveys information solely through the active antenna index, SM expands the concept by transmitting a modulated symbol through the active antenna \cite{DiRenzo2014}. To overcome the requirement that the number of transmit antennas must be a power of two, generalized SM (GSM) was introduced to activate a combination of antennas to transmit the same data symbol \cite{younis2014performance}. Quadrature SM (QSM) doubles SM spatial spectral efficiency by exploiting both real and imaginary signal dimensions. It routes these orthogonal components using a single RF chain coupled with an ideal spatial demultiplexer \cite{mesleh2015quadrature}. Applying these SM schemes to emerging scalable hardware, such as pinching-antenna systems (PASS), is a critical step towards next-generation networks.

PASS consist of waveguides along which pinches can be activated to radiate signals, enabling line-of-sight (LoS) links \cite{xu2025generalized, liu2025pinching}. Recognizing the potential of these architectures, recent literature has begun exploring the application of various SM schemes to PASS. Specifically, SSK has been applied to PASS by utilizing the index of a selected waveguide to convey information bits \cite{ssk}. Additionally, GSM has been integrated into these systems by encoding data through the indices of multiple active pinches \cite{index_modulation}. Furthermore, differential SM enables noncoherent PASS transmission without instantaneous channel state information (CSI) while achieving transmit diversity \cite{diff_sm}. While these studies establish a strong foundation, applying QSM to PASS remains unexplored. Additionally, existing GSM techniques are limited to one-dimensional (1D) spatial indexing, relying solely on the pinch index to convey information, which restricts further spectral efficiency gains. To overcome these limitations, this letter presents the following main contributions.

\begin{itemize}
    \item A 1D-QSM architecture is proposed for PASS, which physically decouples the in-phase and quadrature components across a fixed pair of waveguides, conveying spatial information solely through the pinch indices.
    \item A 2D spatial indexing framework developed for GSM and QSM enhances spectral efficiency by jointly encoding information across active waveguide and pinch indices.
    \item Analytical upper-bound bit error rate (BER) expressions are derived for Rician fading channels to evaluate the performance limits of the proposed schemes.
    \item A computational complexity analysis of the decoder is conducted to demonstrate the practical viability and efficiency of the proposed schemes.
    \item Monte Carlo simulations validate theoretical bounds, and results show that 1D/2D-QSM schemes achieve the lowest decoder complexity. The BER performance depends on spectral efficiency: 2D-QSM excels at lower efficiencies, whereas 2D-GSM performs better at higher ones.
\end{itemize}

\section{System Model}
\label{sec:system_model}

\begin{figure}[htbp]
    \centering
    \includegraphics[width=0.5\linewidth]{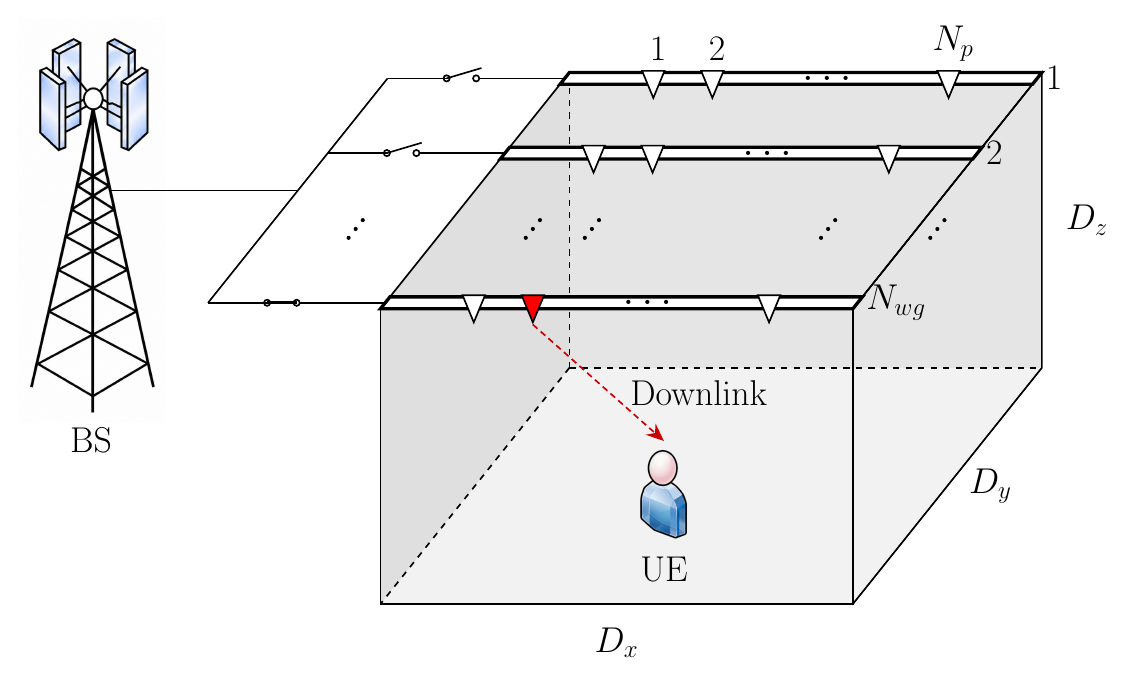}
    \caption{Proposed system model.}
    \label{fig:system_model}
\end{figure}

Consider the downlink MIMO system depicted in Fig.~\ref{fig:system_model}. The system operates within a room defined by an area of $D_x \times D_y$ and a height of $D_z$, where a base station (BS) serves a single user equipment (UE) located within this space. The BS connects via a wired link to a ceiling-mounted array consisting of $N_{wg}$ parallel waveguides. Each waveguide features $N_p$ pinching antennas, resulting in a total of $N_t = N_{wg} N_p$ transmit antennas, where both $N_{wg}$ and $N_p$ are configured to be integer powers of two. Finally, the UE is equipped with an $N_r$-element uniform linear array of receive antennas with a spacing of $\lambda/2$, where $\lambda$ is the free-space wavelength.

\subsection{Channel Model}

The composite channel of the PASS architecture cascades deterministic in-waveguide phase shifts with stochastic free-space propagation. Under an equal-power transmission configuration \cite{index_modulation}, the in-waveguide phase shift for pinch $p$ on waveguide $w$ is
\begin{equation}
    \phi_{w,p} = \exp\left(-\mathrm{j}\frac{2\pi}{\lambda_g} d_{\mathrm{feed},w,p}\right),
\end{equation}
where $d_{\mathrm{feed},w,p}$ is the distance from the RF feed to pinch $p$, $\lambda_g = \frac{\lambda}{n_g}$ is the guided wavelength, and $n_g$ is the effective refractive index. The free-space channel between pinch $p$ on waveguide $w$ and receive antenna $r \in \{0, \dots, N_r-1\}$ follows a Rician fading model \cite{index_modulation, kang2025campass}. Assuming a path loss exponent of 2, the entry at the $r$-th row and $p$-th column of the free-space channel matrix $\mathbf{H}_w^{\mathrm{fs}} \in \mathbb{C}^{N_r \times N_p}$ is
\begin{equation}
    \left[\mathbf{H}_w^{\mathrm{fs}}\right]_{r,p} = \frac{\lambda}{4\pi d_{w,p,r}} \left( \sqrt{\frac{K}{K+1}}\, \bar{h}_{w,p,r} + \sqrt{\frac{1}{K+1}}\, \tilde{h}_{w,p,r} \right),
\end{equation}
where $d_{w,p,r}$ is the Euclidean distance between pinch $p$ and receive antenna $r$, $K$ is the Rician factor, $\bar{h}_{w,p,r} = \exp\left(-\mathrm{j}\frac{2\pi}{\lambda} d_{w,p,r}\right)$ is the deterministic LoS component, and $\tilde{h}_{w,p,r} \sim \mathcal{CN}(0,1)$ represents the non-LoS (NLoS) components, which are assumed to be independent and identically distributed across all pinches and receive antennas. Defining the diagonal in-waveguide phase matrix $\mathbf{\Phi}_w \in \mathbb{C}^{N_p \times N_p}$ as $\mathbf{\Phi}_w = \mathrm{diag}(\phi_{w,0}, \dots, \phi_{w,N_p-1})$, the composite channel sub-matrix for waveguide $w$ is given by $\mathbf{H}_w = \mathbf{H}_w^{\mathrm{fs}} \mathbf{\Phi}_w$. The overall channel matrix $\mathbf{H} \in \mathbb{C}^{N_r \times N_t}$ is formed by concatenating these sub-matrices horizontally, expressed as $\mathbf{H} = [\mathbf{H}_0, \mathbf{H}_1, \dots, \mathbf{H}_{N_{wg}-1}]$.

\subsection{1D Spatial Indexing}
\label{subsec:transmission_schemes}

Let $\eta$ denote the overall spectral efficiency of the system in bits per channel use (bpcu). In SM schemes \cite{DiRenzo2014, index_modulation}, the incoming bit stream is partitioned into spatial bits, which determine the active antenna indices, and constellation bits, which select a symbol $s$ from an $M$-ary quadrature amplitude modulation ($M$-QAM) alphabet $\mathcal{S}$. The complex-valued transmitted symbol is composed of two real-valued pulse amplitude modulation (PAM) symbols representing the in-phase ($s_{\Re}$) and quadrature ($s_{\Im}$) components, such that $s = s_{\Re} + \mathrm{j} s_{\Im}$. The resulting transmission vector is denoted by $\mathbf{x} \in \mathbb{C}^{N_t \times 1}$, such that the received signal $\mathbf{y} \in \mathbb{C}^{N_r \times 1}$ at the user is given by
\begin{equation}
    \mathbf{y} = \sqrt{E_s}\mathbf{H}\mathbf{x} + \mathbf{n},
    \label{eq:received_signal}
\end{equation}
where $E_s$ denotes the average symbol energy, and $\mathbf{n} \sim \mathcal{CN}(\mathbf{0}, \sigma^2\mathbf{I}_{N_r})$ is the additive white Gaussian noise vector, with $\mathbf{I}_{N_r}$ being the $N_r \times N_r$ identity matrix and $\sigma^2$ denoting the noise variance at each receive antenna.

\noindent Existing PASS implementations employ a 1D spatial indexing framework, encoding information exclusively through active pinch indices on a single waveguide for GSM \cite{index_modulation}. However, this configuration introduces challenges when implementing QSM, as a single RF feed forces the entire symbol to be radiated from the active pinches, preventing the independent spatial mapping of $s_{\Re}$ and $s_{\Im}$.

\section{Proposed Solution}
\label{sec:proposed_solution}

To enhance the spectral efficiency of the $N_{wg} \times N_p$ PASS architecture while respecting the single RF-feed limitation, this section proposes a novel dual-waveguide mapping for QSM. Additionally, it introduces a 2D spatial indexing framework for both QSM and GSM, which expands the spatial domain by jointly exploiting the waveguide and pinch antenna indices.

\subsection{Proposed 1D-QSM}
\label{subsec:qsm_pass}

In the proposed 1D-QSM scheme, spatial information is conveyed exclusively through the pinch indices. The system utilizes a fixed pair of in-phase ($w_{\Re}$) and quadrature ($w_{\Im}$) waveguides that carry no additional spatial bits and serve solely to physically decouple $s_{\Re}$ and $s_{\Im}$. The orthogonality of $s_{\Re}$ and $s_{\Im}$ allows them to be independently resolved at the receiver. Specifically, distinct blocks of $\log_2(N_p)$ bits independently select the in-phase pinch index $p_{\Re}$ on $w_{\Re}$ and the quadrature pinch index $p_{\Im}$ on $w_{\Im}$ to transmit $s_{\Re}$ and $s_{\Im}$, respectively. Since $|s_{\Re}|^2 + |s_{\Im}|^2 = |s|^2$ and the square QAM constellation is normalized, the average transmission power is conserved. By fixing the architecture to $N_{wg}=2$ indexed as $w_{\Re} = 0$ and $w_{\Im} = 1$, the proposed 1D-QSM spatial codebook is defined by the set
\begin{equation}
\mathbb{X}_{\text{1D-QSM}} = \left\{ \mathbf{x} \in \mathbb{C}^{N_t \times 1} \;\middle|\;
\begin{array}{l}
    x_{p_{\Re}} = s_{\Re}, \\
    x_{N_p + p_{\Im}} = \mathrm{j} s_{\Im}, \\
    p_{\Re}, p_{\Im} \in \{0, \dots, N_p-1\}, \\
    s \in \mathcal{S}
\end{array}
\right\},
\end{equation}
which yields a total spectral efficiency of
\begin{equation}
\eta_{\text{1D-QSM}} = 2\log_2(N_p) + \log_2(M).
\end{equation}
Consequently, the received signal in \eqref{eq:received_signal} expands to
\begin{equation}
    \mathbf{y} = \sqrt{E_s} \mathbf{h}_{p_{\Re}} s_{\Re} + \mathrm{j} \sqrt{E_s} \mathbf{h}_{N_p + p_{\Im}} s_{\Im} + \mathbf{n},
\end{equation}
where $\mathbf{h}_{p_{\Re}}$ and $\mathbf{h}_{N_p + p_{\Im}}$ are the columns of $\mathbf{H}$ corresponding to the active pinches on $w_{\Re}$ and $w_{\Im}$, respectively.

\subsection{Proposed 2D-QSM}
\label{subsec:2d_indexing}

The proposed 2D-QSM conveys spatial information via both waveguide pair and pinch antenna indices. The $N_{wg}$ available waveguides are partitioned into $N_{\text{pairs}} = 2^{b_{pair}}$ disjoint adjacent pairs, where $b_{pair} = \log_2(N_{wg}/2)$. The architecture groups adjacent waveguides into distinct pairs, designated by the index $\ell \in \{0, \dots, N_{\text{pairs}}-1\}$. The set of valid spatial pairs is formalized as
\begin{equation}
\mathbb{P} = \left\{ \mathcal{P}_{\ell} \;\middle|\; \mathcal{P}_{\ell} = (2\ell, 2\ell+1) \right\}.
\end{equation}
The spatial bit stream is partitioned as follows: the first $b_{pair}$ bits select the pair index $\ell$, activating the waveguide pair $(w_{\Re}, w_{\Im}) = \mathcal{P}_{\ell} \in \mathbb{P}$. The remaining bits select the pinch indices $p_{\Re}$ and $p_{\Im}$ exactly as in 1D-QSM. Therefore, the proposed 2D-QSM spatial codebook is defined by the set
\begin{equation}
\mathbb{X}_{\text{2D-QSM}} = \left\{ \mathbf{x} \in \mathbb{C}^{N_t \times 1} \;\middle|\; 
\begin{array}{l}
    x_{w_{\Re}N_p + p_{\Re}} = s_{\Re}, \\
    x_{w_{\Im}N_p + p_{\Im}} = \mathrm{j} s_{\Im}, \\
    (w_{\Re}, w_{\Im}) = \mathcal{P}_{\ell}, \\
    p_{\Re}, p_{\Im} \in \{0, \dots, N_p-1\}, \\
    s \in \mathcal{S}
\end{array}
\right\},
\end{equation}
which yields a total spectral efficiency of
\begin{equation}
\eta_{\text{2D-QSM}} = \log_2\left(\frac{N_{wg}}{2}\right) + 2\log_2(N_p) + \log_2(M).
\end{equation}
As shown in Fig.~\ref{fig:2d_qsm_example}, consider an input stream of \texttt{0 01 11}. The first bit (\texttt{0}) selects the waveguide pair $\mathcal{P}_0$, activating waveguides $w_{\Re} = 0$ and $w_{\Im} = 1$. The subsequent bit block (\texttt{01}) selects pinch $p_{\Re} = 1$ on $w_{\Re}$ to transmit $s_{\Re}$, and the final bit block (\texttt{11}) selects pinch $p_{\Im} = 3$ on $w_{\Im}$ to transmit $s_{\Im}$.

\subsection{Proposed 2D-GSM}

Let $A_{wg} \ge 1$ and $A_p \ge 1$ denote the number of simultaneously active waveguides and pinches, respectively. All possible combinations for each are lexicographically ordered and truncated to satisfy the power-of-two constraint. This yields the valid spatial combination sets $\mathbb{I}_{wg}$ and $\mathbb{I}_p$, containing $2^{b_{wg}}$ and $2^{b_p}$ elements, respectively, where $b_{wg} = \lfloor\log_2 \binom{N_{wg}}{A_{wg}}\rfloor$ and $b_p = \lfloor\log_2 \binom{N_p}{A_p}\rfloor$.

\begin{figure}[htbp]
    \centering
    \includegraphics[width=0.2\linewidth]{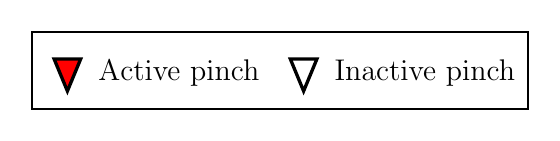}
    \\
    \vspace{-0.5cm}
    \centering
    \subfloat[2D-QSM]{
        \includegraphics[width=0.235\columnwidth]{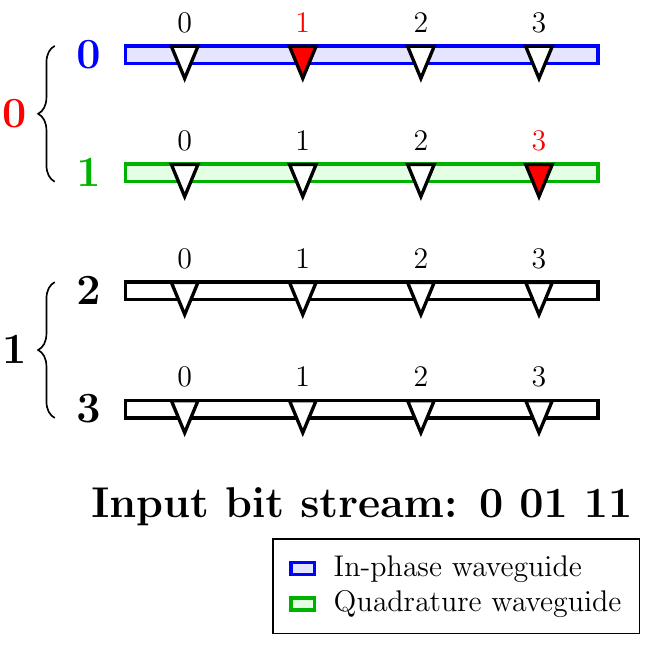}
        \label{fig:2d_qsm_example}
    }
    \subfloat[2D-GSM ($A_{wg}=2$, $A_p=2$)]{
        \includegraphics[width=0.235\columnwidth]{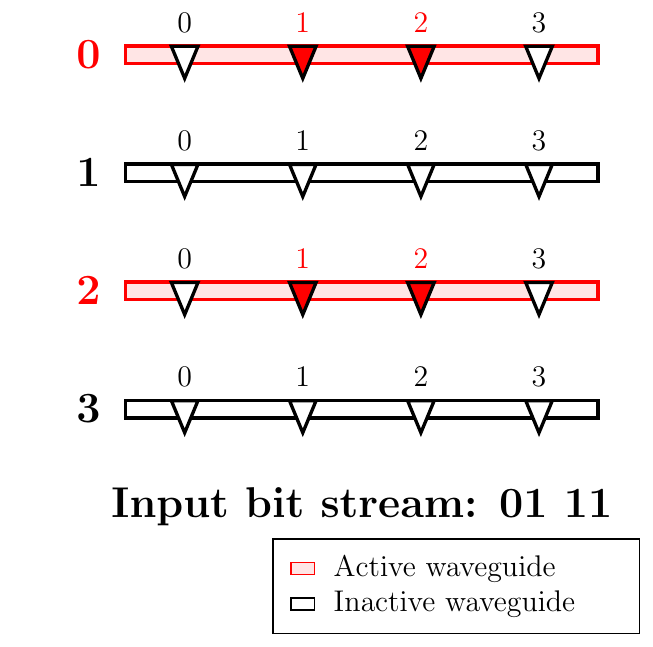}
        \label{fig:2d_gsm_example}
    }
    \caption{2D spatial indexing example for QSM and GSM.}
    \label{fig:2d_examples}
\end{figure}

\noindent The incoming spatial bits index into these master sets to select a specific subset of active waveguides, $\mathcal{I}_{wg} \in \mathbb{I}_{wg}$, and a specific subset of active pinches, $\mathcal{I}_p \in \mathbb{I}_p$. To reduce the decoder search space, the same pinch antenna subset $\mathcal{I}_p$ is activated on every selected waveguide $w \in \mathcal{I}_{wg}$. Finally, the transmit power is equally distributed among the total number of active antennas, $N_a = A_{wg} A_p$, ensuring the overall transmission power remains constant. Note that setting $N_a = 1$ reduces this architecture to 2D-SM. The proposed 2D-GSM codebook is given by
\begin{equation}
\mathbb{X}_{\text{2D-GSM}} = \left\{ \mathbf{x} \in \mathbb{C}^{N_t \times 1} \;\middle|\;
\begin{array}{l}
x_{w N_p + p} = \frac{s}{\sqrt{N_a}}, \\
\forall w \in \mathcal{I}_{wg}, \forall p \in \mathcal{I}_p, \\
\mathcal{I}_{wg} \in \mathbb{I}_{wg}, \mathcal{I}_p \in \mathbb{I}_p, s \in \mathcal{S}
\end{array}
\right\},
\end{equation}
with a total spectral efficiency of
\begin{equation}
\eta_{\text{2D-GSM}} = \left\lfloor\log_2 \binom{N_{wg}}{A_{wg}}\right\rfloor + \left\lfloor\log_2 \binom{N_p}{A_p}\right\rfloor + \log_2(M).
\end{equation}
As illustrated in Table~\ref{tab:spatial_mapping} and Fig.~\ref{fig:2d_gsm_example}, consider an input stream of \texttt{01 11}. The first bit block (\texttt{01}) selects the active waveguide index set $\mathcal{I}_{wg} = \{0, 2\}$, while the second bit block (\texttt{11}) selects the active pinch index set $\mathcal{I}_p = \{1, 2\}$, activating those specific pinches across all the selected waveguides.

\vspace{-0.5cm}
\begin{table}[htbp]
    \centering
    \caption{Spatial Bit Mapping and Combinatorial Truncation for 2D-GSM ($N_{wg}=4, N_p=4, A_{wg}=2, A_p=2$)}
    \label{tab:spatial_mapping}
    \begin{tabular}{@{}c c@{}}
        \toprule
        \textbf{Spatial Bits} & \textbf{Active Elements} \\
        ($b_1 b_0$) & $\mathcal{I} \in \mathbb{I}$ \\
        \midrule
        00 & $\{0, 1\}$ \\
        01 & $\{0, 2\}$ \\
        10 & $\{0, 3\}$ \\
        11 & $\{1, 2\}$ \\
        \midrule
        \textit{Combinations discarded} & $\{1, 3\}$ \\
        \textit{to satisfy power-of-two constraint} & $\{2, 3\}$ \\
        \bottomrule
    \end{tabular}
\end{table}

\begin{table*}[t]
\centering
\caption{Computational Complexity of the ML Detector for Scheme $\Omega$}
\label{tab:complexity}
\renewcommand{\arraystretch}{1.3}
\setlength{\tabcolsep}{4.5pt}
\begin{tabular}{l c c c c c}
\toprule
\hspace{0.3cm} $\Omega$ 
& $N_{\text{sym}}^{\Omega}$ 
& $C_{RM}^{\Omega}$ 
& $C_{RA}^{\Omega}$ 
& $C^{\Omega}$ 
& \textbf{Asymptotic Complexity} \\
\midrule

1D-QSM 
& $2 N_p \sqrt{M}$ 
& $4 N_r N_{\text{sym}}^{\Omega}$ 
& $(4 N_r - 1) N_{\text{sym}}^{\Omega}$ 
& $(8 N_r - 1) N_{\text{sym}}^{\Omega}$ 
& $\mathcal{O}\!\left(N_r N_p \sqrt{M}\right)$ \\

2D-QSM 
& $N_{wg} N_p \sqrt{M}$ 
& $4 N_r N_{\text{sym}}^{\Omega}$ 
& $(4 N_r - 1) N_{\text{sym}}^{\Omega}$ 
& $(8 N_r - 1) N_{\text{sym}}^{\Omega}$ 
& $\mathcal{O}\!\left(N_r N_{wg} N_p \sqrt{M}\right)$ \\

2D-GSM 
& $2^{\left\lfloor \log_2 \binom{N_{wg}}{A_{wg}} \right\rfloor}
   2^{\left\lfloor \log_2 \binom{N_p}{A_p} \right\rfloor} M$ 
& $6 N_r N_{\text{sym}}^{\Omega}$ 
& $(2 N_r N_a + 4 N_r - 1) N_{\text{sym}}^{\Omega}$ 
& $(2 N_r N_a + 10 N_r - 1) N_{\text{sym}}^{\Omega}$ 
& $\mathcal{O}\!\left(N_r M \binom{N_{wg}}{A_{wg}} \binom{N_p}{A_p}\right)$ \\

\bottomrule
\end{tabular}
\end{table*}

\section{Decoder Performance and Complexity Analysis}
Assuming perfect CSI at the receiver, the optimal maximum likelihood (ML) detector jointly estimates the active spatial indices and the transmitted symbol via an exhaustive search over all valid transmission vectors in the defined spatial codebook $\mathbb{X}_{\Omega}$, where $\Omega \in \{\text{1D-QSM}, \text{2D-QSM}, \text{2D-GSM}\}$. The estimated vector $\hat{\mathbf{x}}_{\text{ML}}^{\Omega}$ is given by
\begin{equation}
    \hat{\mathbf{x}}_{\text{ML}}^{\Omega} = \arg \min_{\mathbf{x}^{\Omega} \in \mathbb{X}_{\Omega}} \left\| \mathbf{y} - \sqrt{E_s}\mathbf{H}\mathbf{x}^{\Omega} \right\|^2.
\end{equation}

\subsection{Upper-bound BER Analysis}
The average bit error probability via union bounding is upper-bounded as \cite{simon2005digital}
\begin{equation}
    P_{b}^{\Omega} \le \frac{1}{2^{\eta_{\Omega}}} \sum_{i=1}^{2^{\eta_{\Omega}}} \sum_{\substack{j = 1 \\ j \neq i}}^{2^{\eta_{\Omega}}} P(\mathbf{x}_{i}^{\Omega} \rightarrow \mathbf{x}_{j}^{\Omega}) \frac{e_{i,j}^{\Omega}}{\eta_{\Omega}},
\label{eq:bit_error_bound}
\end{equation}
where $P(\mathbf{x}_{i}^{\Omega} \rightarrow \mathbf{x}_{j}^{\Omega})$ represents the pair-wise error probability (PEP), and $e_{i,j}^{\Omega}$ represents the number of bit errors resulting from each PEP event under scheme $\Omega$. Let $\gamma_{i,j}^{\Omega} = \left\| \mathbf{H}\boldsymbol{\Delta}_{i,j}^{\Omega} \right\|^{2}$, where $\boldsymbol{\Delta}_{i,j}^{\Omega} = \mathbf{x}_{i}^{\Omega} - \mathbf{x}_{j}^{\Omega}$ represents the global symbol difference vector between the transmitted and erroneously detected spatial vectors. Averaging the conditional PEP over the fading distribution of $\gamma_{i,j}^{\Omega}$ yields the unconditional PEP, given by
\begin{equation}
    P(\mathbf{x}_{i}^{\Omega} \rightarrow \mathbf{x}_{j}^{\Omega}) = \mathbb{E}_{\gamma_{i,j}^{\Omega}} \left[ Q\left( \sqrt{\frac{E_s \gamma_{i,j}^{\Omega}}{2N_{0}}} \right) \right].
\end{equation}
Using Craig's exact integral representation for the Q-function \cite{simon2005digital}, the unconditional PEP can be expressed via the moment generating function, $\mathcal{M}_{\gamma_{i,j}^{\Omega}}(t) = \mathbb{E}[e^{\gamma_{i,j}^{\Omega}t}]$, yielding
\begin{equation}
    P(\mathbf{x}_{i}^{\Omega} \rightarrow \mathbf{x}_{j}^{\Omega}) = \frac{1}{\pi} \int_{0}^{\pi/2} \mathcal{M}_{\gamma_{i,j}^{\Omega}}\left( -\frac{E_s}{4N_0 \sin^2\theta} \right) d\theta.
\end{equation}
Approximating this integral using Gauss-Chebyshev quadrature with $N_{gc}$ quadrature nodes \cite{gauss_chebyshev} gives
\begin{equation}
    P(\mathbf{x}_{i}^{\Omega} \rightarrow \mathbf{x}_{j}^{\Omega}) \approx \frac{1}{2N_{gc}} \sum_{m=1}^{N_{gc}} \mathcal{M}_{\gamma_{i,j}^{\Omega}}\left( -c_m \right),
\end{equation}
where $c_m = \frac{E_s}{4N_{0} \sin^2 \theta_m}$ and $\theta_m = \frac{(2m-1)\pi}{4N_{gc}}$. Applying column-wise vectorization to the composite channel matrix yields $\mathbf{v} \triangleq \mathrm{vec}(\mathbf{H})$, such that the received difference vector becomes $\mathrm{vec}(\mathbf{H}\boldsymbol{\Delta}_{i,j}^{\Omega}) = ((\boldsymbol{\Delta}_{i,j}^{\Omega})^T \otimes \mathbf{I}_{N_r})\mathbf{v}$. Thus, $\gamma_{i,j}^{\Omega}$ can be expressed as a Hermitian quadratic form \cite{index_modulation}, given by
\begin{equation}
    \gamma_{i,j}^{\Omega} = \mathbf{v}^{H} \mathbf{A}_{i,j}^{\Omega} \mathbf{v},
\end{equation}
where $\mathbf{A}_{i,j}^{\Omega} = ((\boldsymbol{\Delta}_{i,j}^{\Omega})^* (\boldsymbol{\Delta}_{i,j}^{\Omega})^T) \otimes \mathbf{I}_{N_r}$. Since the NLoS components are modeled as complex Gaussian, $\mathbf{v}$ is a complex Gaussian random vector characterized by its mean vector $\overline{\mathbf{v}} = \mathbb{E}\{\mathbf{v}\}$ and its covariance matrix $\mathbf{C}_{\mathbf{v}} = \mathbb{E}\left\{(\mathbf{v} - \overline{\mathbf{v}})(\mathbf{v} - \overline{\mathbf{v}})^H\right\}$. Letting $\mathbf{\Psi}_{i,j,m}^{\Omega} = \mathbf{I}_{N_r N_t} + c_m\mathbf{C}_{\mathbf{v}}\mathbf{A}_{i,j}^{\Omega}$, the unconditional PEP expression becomes
\begin{equation}
    P(\mathbf{x}_{i}^{\Omega} \rightarrow \mathbf{x}_{j}^{\Omega}) \approx \frac{1}{2N_{gc}} \sum_{m=1}^{N_{gc}} \frac{\exp\left(-c_m \overline{\mathbf{v}}^{H}\mathbf{A}_{i,j}^{\Omega}(\mathbf{\Psi}_{i,j,m}^{\Omega})^{-1}\overline{\mathbf{v}}\right)}{\text{det}\left(\mathbf{\Psi}_{i,j,m}^{\Omega}\right)}.
    \label{eq:pep}
\end{equation}
While substituting \eqref{eq:pep} into \eqref{eq:bit_error_bound} establishes the theoretical upper bound for the BER, the practical viability of these schemes depends on their detection computational complexity.

\subsection{Complexity Analysis}
The complexity of the optimal ML detector is defined by the real multiplications ($RM$) and real additions ($RA$) required per hypothesis, as well as the search space size ($N_{\text{sym}}^\Omega$). By exploiting the sparsity of $\mathbf{x}$, SM schemes extract only the active channel columns to avoid full $N_t$-dimensional matrix multiplications. This reduces the total real operations ($C^{\Omega}$), calculated as the sum of total $RM$ ($C_{\text{RM}}^{\Omega}$) and $RA$ ($C_{\text{RA}}^{\Omega}$). The asymptotic complexity $\mathcal{O}(\cdot)$ is derived from $C^{\Omega}$ by dropping constant arithmetic scalars and omitting the floor operations to highlight the scaling of computational loads. Table \ref{tab:complexity} details these operations for each proposed scheme. 

\noindent Due to the orthogonality of $s_{\Re}$ and $s_{\Im}$, ML detection in 1D/2D-QSM decouples into two independent searches \cite{alnahhal2017qsm}. By separately estimating the active pinch index and the $\sqrt{M}$-PAM for the in-phase and quadrature components, the computational complexity is significantly reduced compared to a full complex $M$-QAM evaluation. While 1D-QSM limits its spatial search to a fixed waveguide pair, 2D-QSM independently evaluates the optimal pinch across all possible $w_{\Re}$ and $w_{\Im}$. For each receive antenna, evaluating the ML metric takes 4 $RM$ (scalar-complex multiplication and squaring differences) and 3 $RA$ (computing differences and summing squares). Scaling across the $N_r$ antennas and adding the $(N_r - 1)$ $RA$ needed to sum the metrics across the array yields exactly $4N_r$ $RM$ and $(4N_r - 1)$ $RA$ per hypothesis. Conversely, 2D-GSM aggregates the channel columns of the $N_a$ active antennas \cite{younis2014performance}, requiring $2(N_a - 1)$ $RA$ per receive antenna. Evaluating the metric against the QAM symbol requires 6 $RM$ (complex multiplication and squaring differences) and 5 $RA$ (complex multiplication, differences, and summing squares). This scales to $6N_r$ $RM$ across the array. Including the $(N_r - 1)$ operations to sum the metrics across all antennas, the overall complexity totals $(2N_rN_a + 4N_r - 1)$ $RA$ per hypothesis.

\section{Simulation Results}

\begin{figure*}[t]
    \centering
    \subfloat[$\eta = 8$ bpcu and $N_t = 16$.]{
        \includegraphics[width=0.3\textwidth]{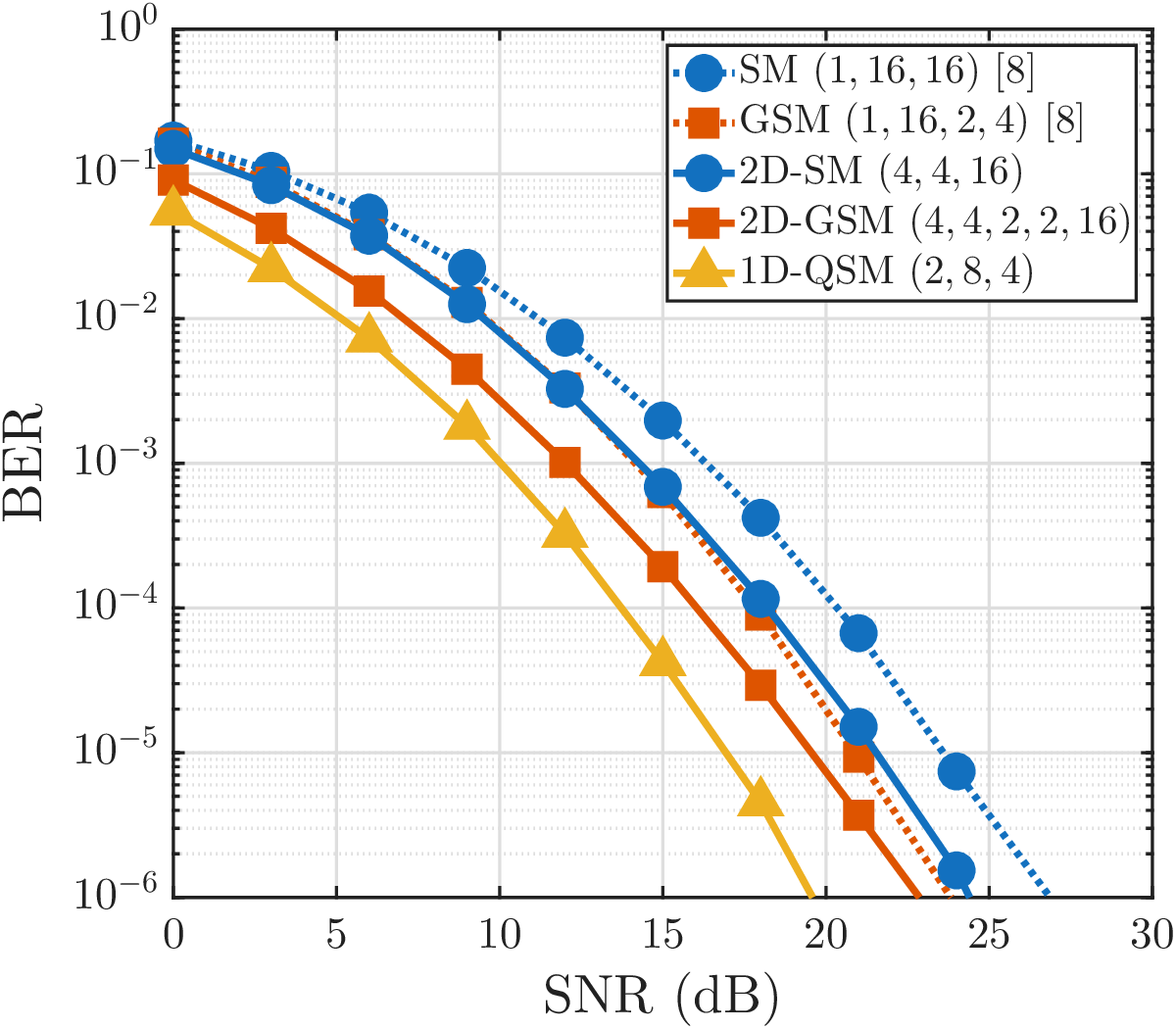}
        \label{fig:1d_2d}
    }
    \hfill
    \subfloat[$\eta = 9$ bpcu, $N_{wg} = 4$ and $N_p = 8$.]{
        \includegraphics[width=0.3\textwidth]{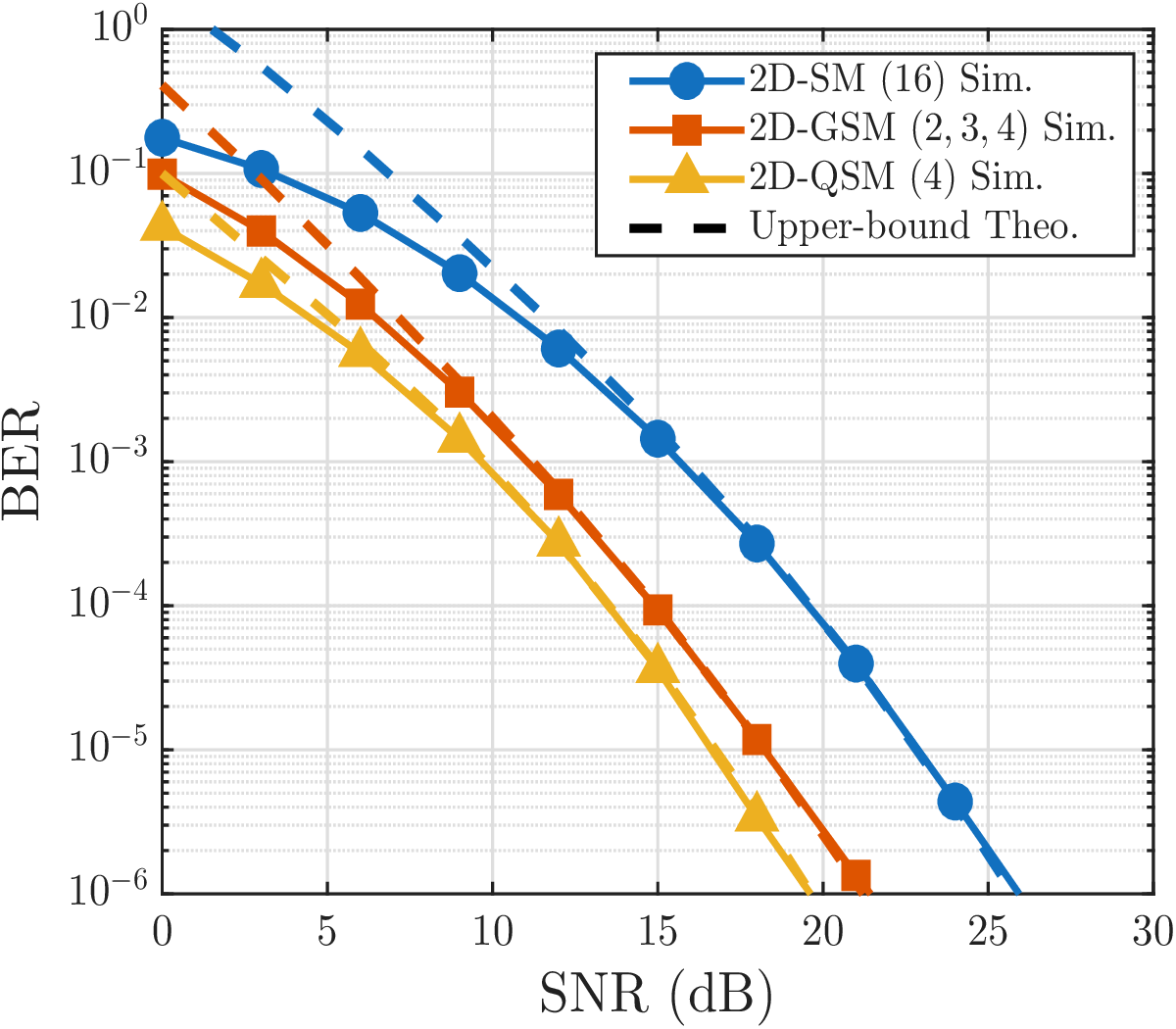}
        \label{fig:gsm_qsm}
    }
    \hfill
    \subfloat[$\eta = 13$ bpcu, $N_{wg} = 4$ and $N_p = 16$.]{
        \includegraphics[width=0.3\textwidth]{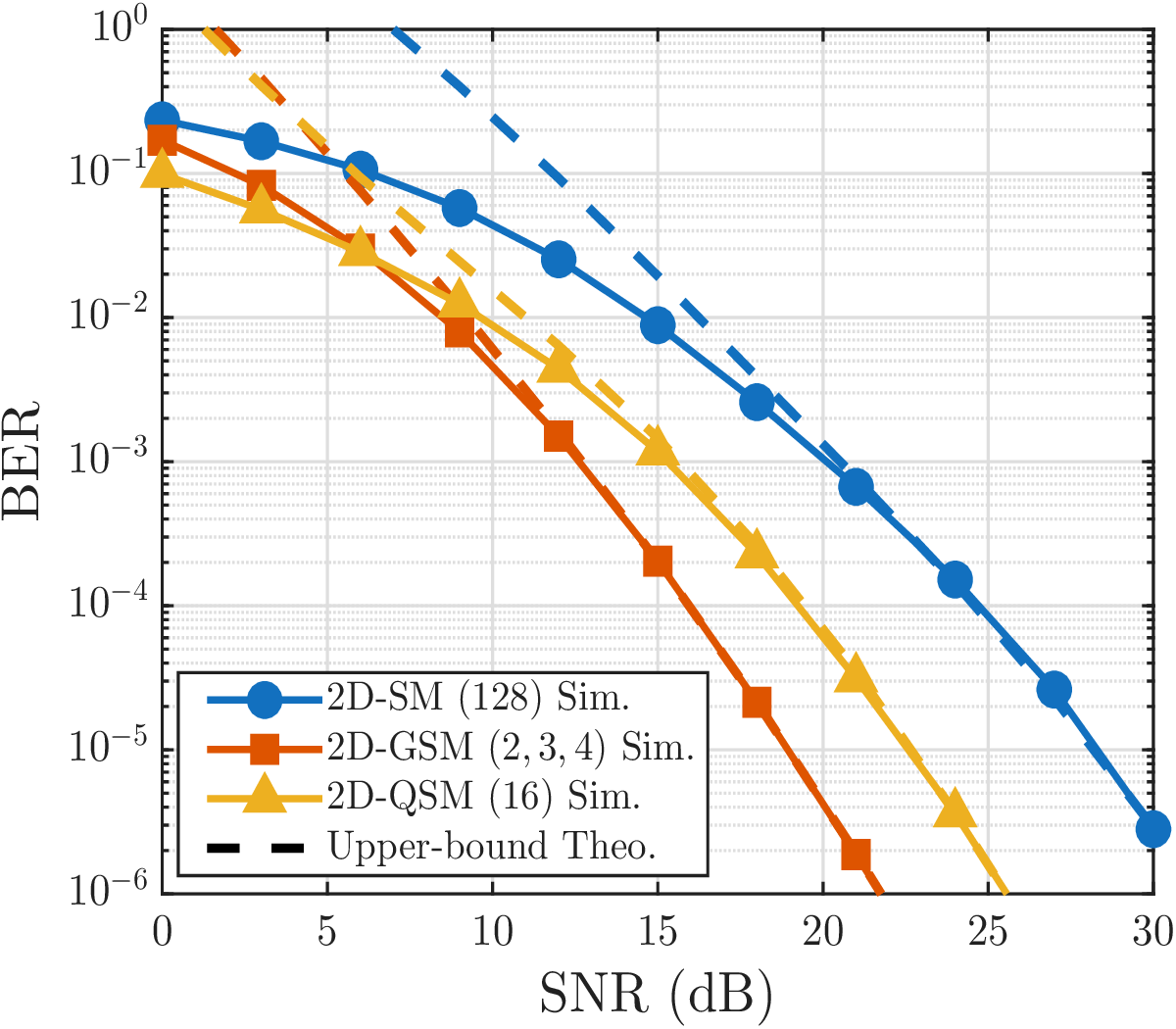}
        \label{fig:gsm_qsm2}
    }
    
    \caption{BER comparison of the proposed schemes. Notation in (a) is $(N_{wg}, N_p, M)$ for SM \cite{index_modulation}, 2D-SM and 1D-QSM, $(N_{wg}, N_p, A_p, M)$ for GSM \cite{index_modulation}, and $(N_{wg}, N_p, A_{wg}, A_p, M)$ for 2D-GSM. Notation in (b) and (c) is $(M)$ for 2D-SM and 2D-QSM, and $(A_{wg}, A_p, M)$ for 2D-GSM.}
    \label{fig:combined_simulation_results}

    \vspace{-0.1cm}
\end{figure*}

The BER of the proposed schemes is evaluated against the signal-to-noise ratio (SNR) using Monte Carlo simulations for a 5 GHz system and $n_g = 1.4$ in an indoor Rician fading environment with $K = 10$ dB. Within a room of dimensions $D_x = D_y = 20$ m and $D_z = 3$ m, a stationary UE with $N_r = 4$ antennas is uniformly distributed across the $x$-$y$ plane at a fixed height of 1 m and assigned a random azimuth orientation for each realization. Theoretical upper bounds are validated using $N_{gc} = 20$.

Fig.~\ref{fig:1d_2d} compares the BER of the proposed 1D-QSM, 2D-SM, and 2D-GSM schemes against 1D baselines \cite{index_modulation} under fixed $\eta$ and $N_t$. The 2D-SM and 2D-GSM configurations outperform their 1D counterparts, while GSM maintains its superiority over SM in both 1D and 2D variants. Overall, 1D-QSM achieves the best performance among all evaluated schemes, delivering SNR gains of approximately 6 dB over SM \cite{index_modulation} and 3 dB over GSM \cite{index_modulation} at a BER of $10^{-5}$. This superiority is driven by the orthogonality of the decoupled symbol components, which eliminates mutual interference and drastically shrinks $N_{\text{sym}}^{\text{1D-QSM}}$. Consequently, the computational burden of 1D-QSM is reduced to $C^{\text{1D-QSM}} = 992$, representing a 91.8\% reduction compared to SM \cite{index_modulation} ($C^{\text{SM}} = 12{,}032$) and a 93.0\% decrease against GSM \cite{index_modulation} ($C^{\text{GSM}} = 14{,}080$). Finally, the complexity of 2D-SM matches that of SM \cite{index_modulation} since $N_t$ and $M$ are fixed, whereas the 2D-GSM complexity increases by 29\% compared to GSM \cite{index_modulation} due to the increased $A_{wg}$.

Figs.~\ref{fig:gsm_qsm} and \ref{fig:gsm_qsm2} validate the theoretical upper bounds against the simulated BER curves for the proposed 2D schemes. At $\eta = 9$ bpcu and a BER of $10^{-5}$ (Fig.~\ref{fig:gsm_qsm}), 2D-QSM yields SNR gains of 5.3 dB over 2D-SM and 1.4 dB over 2D-GSM. This is because 2D spatial indexing does not affect the orthogonality of $s_{\Re}$ and $s_{\Im}$, resulting in superior BER performance for 2D-QSM. Furthermore, 2D-QSM requires only $C^{\text{2D-QSM}} = 1{,}984$, reducing complexity by 91.8\% compared to 2D-SM ($C^{\text{2D-SM}} = 24{,}064$) and 95.5\% compared to 2D-GSM ($C^{\text{2D-GSM}} = 44{,}544$). However, as the spectral efficiency increases to $\eta = 13$ bpcu (Fig.~\ref{fig:gsm_qsm2}), the BER performance of 2D-GSM surpasses that of 2D-QSM. At a target BER of $10^{-5}$, 2D-GSM yields SNR gains of 10 dB and 4 dB over 2D-SM and 2D-QSM, respectively. This is because 2D-GSM achieves higher spatial spectral efficiency. Specifically, it gains one bit in the pinch-index dimension through a larger $A_p$, and one bit in the waveguide-index dimension because 2D-QSM uses waveguide pairs to separate the symbol components, leaving only $N_{wg}/2$ indices available. Despite this, 2D-QSM maintains a substantial computational advantage. Requiring only $C^{\text{2D-QSM}} = 7{,}936$, 2D-QSM reduces computational complexity by 97.9\% relative to 2D-SM ($C^{\text{2D-SM}} = 385{,}024$) and by 98.9\% compared to 2D-GSM ($C^{\text{2D-GSM}} = 712{,}704$).

\section{Conclusion}

In this letter, a multi-waveguide PASS architecture was leveraged to implement three novel SM schemes: 1D-QSM, 2D-QSM, and 2D-GSM. By evaluating these schemes in terms of $\eta$, BER, and decoder computational complexity, the following three key conclusions are drawn. First, 2D-SM and 2D-GSM schemes outperform their 1D counterparts. Second, by exploiting the orthogonality of in-phase and quadrature components, 1D/2D-QSM achieve the lowest complexity, demonstrating up to a 98.9\% reduction in total real operations relative to 2D-GSM. Third, performance depends on the target $\eta$. At a BER of $10^{-5}$, 1D/2D-QSM deliver superior performance at lower spectral efficiencies, providing a 6 dB SNR gain over 1D-SM at $\eta = 8$ bpcu; conversely, 2D-GSM offers better performance at high $\eta$, yielding a 10 dB gain over 2D-SM at $\eta = 13$ bpcu. Ultimately, the proposed schemes offer a flexible tradeoff between BER performance and complexity for a given $\eta$. Future work will investigate multi-user extensions and reduced-complexity detection for large-scale arrays. 

\bibliographystyle{IEEEtran}
\bibliography{references}

\end{document}